\documentclass[lineno]{jfm}

\usepackage{graphicx}
\usepackage{newtxtext}
\usepackage{newtxmath}
\usepackage{natbib}
\usepackage{hyperref}
\hypersetup{
    colorlinks = true,
    urlcolor   = blue,
    citecolor  = black,
}
\newcommand{\RomanNumeralCaps}[1]
\linenumbers

\usepackage{amssymb}
\usepackage{amsmath}
 \usepackage{dsfont}
 \usepackage{tikz}
 \usetikzlibrary{shapes, arrows}
 \usetikzlibrary{arrows.meta,shadows,positioning}
 \usepackage[normalem]{ulem}
\usepackage{mathrsfs,amsfonts}
\usepackage{graphics,color}
 \usepackage{epsfig}
 \usepackage{bm}
 \usepackage[]{subfig}
 \usepackage{color}
 \usepackage[utf8]{inputenc}
 \usepackage[T1]{fontenc}
 \usepackage{mathtools}
 \usepackage{nicefrac}
 \usepackage{csquotes}
 \usepackage{float}

\newcommand*\diff{\mathop{}\!\mathrm{d}}

\newcommand{\rom}[1]{\uppercase\expandafter{\romannumeral #1\relax}}

\DeclareMathOperator{\tr}{\mathrm{tr}}

\DeclareMathOperator{\dev}{\mathrm{dev}}

\begin{document}
%\nocite{*}
\title{A viscoelastic phase-field model for iceberg calving}
\author{Jakub Stocek, Robert J. Arthern, Oliver J. Marsh}
\date{...}
\maketitle
\abstract{Iceberg calving accounts for around half of the ice lost annually from Antarctica, but realistic representation of fracture and calving in large-scale ice sheet models remains a major unsolved problem in glaciology. We present a new phase-field viscoelastic model for fracture that simulates the slow deformation of ice and the distribution and evolution of cracks. Cracks nucleate and propagate in response to the evolving stress field, and are influenced by water pressure below sea level. The model incorporates nonlinear-viscous rheology, linear-elastic rheology, and a phase-field variational formulation, which allows simulation of complex fracture phenomena. We show that this approach is capable of simulating the physical process of calving. Numerical experiments supported by a simplified model suggest that calving rate will scale with the fourth power of ice thickness for a floating ice front that has no variation across flow. The equations make no assumptions about the style of calving, so they would also simulate numerous more realistic settings in Antarctica for which material parameters and three-dimensional effects can be expected to influence the calving rate.}
\vskip 0.8cm

%%%%%%%%%%%%%%%%%%%%%%%%%%%%%%%%%%%%%%%%%%%%%%%%%%%%%%%%%%%%%%%%%%%%%%%%%%%

\section{Introduction}\label{sec:introduction}

Calving of tabular icebergs from ice shelves accounts for approximately half the ice lost from Antarctica each year \citep{greene_antarctic_2022}. Failure of calving ice cliffs is also hypothesised to make the Antarctic Ice Sheet vulnerable to an instability that could drive rapid sea level rise \citep{deconto_contribution_2016,pattyn_greenland_2018}. Despite its central importance to the problem of predicting global sea level, the fracture process that drives calving, comprising both initiation and evolution of cracks, is complex and not yet well understood. Rather than simulating the physical process of fracture in detail, most models of calving and cliff retreat rates have been based on heuristic arguments or limited observations scaled up to the whole of Antarctica \citep{deconto_contribution_2016}. There is no guarantee that current methods will accurately capture the sea level contributions when conditions deviate from present-day observations, so a focus on physically-based modelling of fracture in glaciological settings is needed.\\

Glacial ice can be modelled as a viscoelastic material with Maxwell rheology and nonlinear viscosity dependent on the strain rate \citep{glen_creep_1955}. It is more than twenty years since \citet{meier_1997} emphasised that ``iceberg calving is largely a problem in fracture mechanics coupled to ice dynamics". This remark was later highlighted by \citep{benn_calving_2007} as crucial to understanding calving. Despite this, there have been relatively few attempts to engage fully with the implications of this statement. Since \citet{griffith_vi_nodate} it has been clear that brittle fracture inherently involves the transfer of stored elastic energy into surface energy, yet most large-scale ice sheet models neglect elasticity in their formulation, on the assumption that timescales of interest far exceed the Maxwell timescale at which elastic behaviour transitions to viscous behaviour. Similarly, the literature in fracture mechanics tends to concentrate on the elastic problem in isolation, because the timescale under which many engineering samples undergo catastrophic failure can be considered short enough to neglect viscous deformation. To accept Meier's statement at face value is to acknowledge that calving is a coupled viscoelastic fracture problem, and requires a system of equations that involves ice viscosity, elasticity and brittle fracture. This is the framework that we adopt here. Following \citep{miehe_phase_2010} we use phase-field approach to simulate the brittle failure of ice. \\

%%2. Lit review
Previously, fracture and crevasse propagation in ice have been addressed using a combination of analytical and computational techniques. Approximations for crevasse depths in idealised geometries have been described using a zero-stress model advanced by \citet{nye_comments_1955}, \citet{meier_mechanics_1958}, and \citet{weertman_can_1973}. Under this paradigm, crevasses are assumed to form under any tensile stress, no matter how small, but to stop propagating at the `Nye depth', where the compressive component of stress generated by the weight of overlying ice exceeds the tensile component, generated by longitudinal stretching. Further extensions have been made to incorporate water filled crevasses \citep{benn_calving_2007}.\\

Computational modelling of damage evolution in ice shelves and ice sheets has been investigated in \citep{bassis_upper_2012,clerc_marine_2019,lhermitte_damage_2020,mosbeux_effect_2023}. Models based on linear elastic fracture mechanics have been used to estimate crevasse depths on the assumption that ice behaves elastically on short time scales \citep{van_der_veen_fracture_1998,lipovsky_ice_2020,zarrinderakht_effect_2022}. These models can make useful predictions, but are most applicable to idealised geometries with simple boundary conditions, and explicitly specified initial flaws. The heuristic criteria needed to predict the onset of crack nucleation can make such models difficult to parameterise. Implementing numerical methods that represent bifurcation and coalescence of cracks also remains a challenge within the framework of linear elastic fracture mechanics. Together, these factors have discouraged widespread use of linear elastic fracture mechanics for large-scale ice sheet modelling. \\

%%3. Phase field works
One can overcome many of the drawbacks of linear elastic fracture mechanics with diffusive crack modelling. This uses variational approaches that are based on energy minimisation \citep{francfort_revisiting_1998}. {Introducing} a phase field for fracture, one effectively has a variable that interpolates between the solid material and fracture induced voids in a sufficiently smooth manner. This alleviates issues related to the complex crack topology. The crucial difference for numerical treatment is the fact that the regularised problem does not require an explicit treatment of the crack configuration. All computations can be executed on a fixed mesh using standard finite-element techniques.
Further, regularised functionals based on phase-field formulations are $\Gamma$--convergent  to the sharp crack topology functionals for vanishing length-scale regularisation parameter \citep{ambrosio_approximation_1990}. In practice, this means that the correct transfers of energy between stored elastic energy and surface energy are approached as the regularistaion lengthscale decreases.\\

Recently, phase-field models for fracture have gained a large following due to their ability to predict complex cracking phenomena such as crack branching and coalescence, or crack nucleation. A considerable amount of research has been focused on brittle fracture in elastic solids \citep{bourdin_numerical_2000,miehe_phase_2010,miehe_thermodynamically_2010}. New phase-field models have been developed for dynamic fracturing, fluid-driven fracture propagation \citep{mikelic_quasi-static_2015,mikelic_phase-field_2015,mikelic_phase-field_2015-1}, as well as thermo-, visco-, elasto-, plastic materials \citep{miehe_multi-field_2011,miehe_mixed_2012,miehe_phase-field_2016,miehe_phase-field_2017,miehe_phase_2014}. \\

% What do we do?
In this work we present a phase-field formulation of fracture for Maxwell viscoelastic materials. This model is capable of capturing the creep of glacial ice as well as an instantaneous elastic deformation. Phase-field approaches have been used to predict hydrofracture in compressible elastic glacial ice \citep{sun_poro-damage_2021} as well as incompressible viscous materials \citep{clayton_stress-based_2022}. A model for viscoelastic materials with phase-field fracture has been also discussed in \citep{shen_fracture_2019}. They focus on short-term evolution of the fracture network in ice. In contrast, our goal is to describe a model capable of long time evolution, crucial for modelling calving from ice shelves.  \\

%%4. Outline
This article is organised as follows: Section~\ref{sec:model_description} discusses the viscoelastic phase-field fracture model for ice shelf dynamics. For convenience, we divide the presentation into three subsections. Subsection~\ref{ssec:geometry} introduces notation for the model domain and Subsection~\ref{ssec:ice_rheology} presents ice rheology and the underlying equations.  In Subsection~\ref{ssec:phase_field} we review relevant information from phase-field modelling of fracture, incorporate them into the rheology, and present a non-dimensional form of the full system. In Section~\ref{sec:implementation} we discuss implementation details. Numerical experiments are presented in Section~\ref{sec:numerics}, accompanied with discussion of the results in Section~\ref{sec:discussion}. We present concluding remarks in Section~\ref{sec:conclusions}. The article is accompanied by appendices and supplementary materials that address technical details.

%%%INTRO END%%%
\section{Model description}\label{sec:model_description}
In this section, we describe the relevant equations governing the evolution of ice sheets and fracture phase-field equations. We begin by describing model geometry and its associated notation. Next we discuss the rheological properties of ice flow and introduce their energy storage and dissipation potential functions. We then briefly review fracture phase-field equations and include the additional variables into the system of equations that govern ice sheet evolution. We incorporate a hydrostatic fluid pressure condition into the cracks. Finally, we present the nondimensionalised system of equations that govern the evolution of the viscoelastic material and the fracture phase-field equations.

\subsection{Geometry}\label{ssec:geometry}

\begin{figure}[!ht]
\centering
\includegraphics[width=0.7\textwidth]{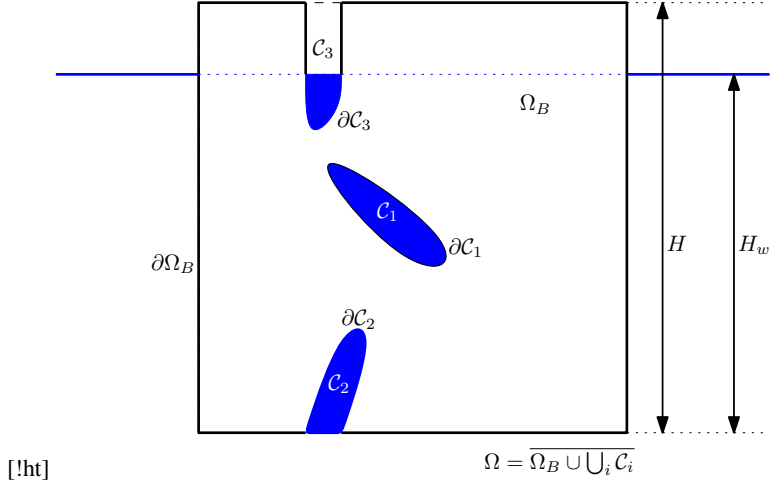}
\caption{Illustration of geometry and associated notation.} 
\label{f:domain2}
\end{figure}

Throughout this article we consider an intact body $\Omega_B$ with inclusions $\mathcal{C}_i$. We introduce an extended domain $\Omega = \overline{ \Omega_B \cup \bigcup_i \mathcal{C}_i}$ which combines both the intact body $\Omega_{B}$ and the inclusions $\mathcal{C}_i$. Denote the boundary of $\mathcal{C}_i$ as $\partial \mathcal{C}_i = \partial_E \mathcal{C}_i \cup \partial_I \mathcal{C}_i$, where we decompose the boundary into exterior and interior parts with respect to $\Omega$, respectively. We further denote boundaries of $\Omega_B$ and $\Omega$ by $\partial \Omega_B = \bigcup_i \partial_I \mathcal{C}_i \cup \partial_E \Omega_B$ and $\partial \Omega = \bigcup_i \partial_E \mathcal{C}_i \cup \partial_E \Omega_B$, respectively.\\

We further decompose the boundary $\partial \Omega_B = \partial_D \Omega_B \cup \partial_N \Omega_B$ according to the boundary conditions. Here $\partial_D \Omega_B, \ \partial_N \Omega_B$ correspond respectively to Dirichlet and Neumann, boundary conditions. Similarly we decompose the boundary of the extended domain $\partial \Omega = \partial_D \Omega \cup \partial_N \Omega$.\\

An example of a domain with three inclusions is illustrated in Figure~\ref{f:domain2}. Here,
$\partial_E \Omega_B$ is the bold part of the boundary, $\partial_E \mathcal{C}_i$ is the dashed part of the boundary, and $\partial_I \mathcal{C}_i$ is the solid fine line of the boundary. Even though the inclusions in Figure~\ref{f:domain2} are represented as having a volume, in case of sharp cracks we will have to treat them as a lower dimensional object. In that case we treat $\mathcal{C}_i$ and $\partial \mathcal{C}_i$ interchangeably. This detail is alleviated by the phase field formulation. \\

We also indicate the height of the domain as $H$ and the height of the water level measured from the base as $H_w$. In the case of a freely floating iceberg we note that $\partial_D \Omega_B = \emptyset$.

\subsection{Ice Rheology}\label{ssec:ice_rheology}
First we will focus only on the behaviour of the intact material $\Omega_{B}$ and neglect any inclusions in the material. We aim to discuss the rheological relations governing the evolution of polycrystalline ice. 
Behaviour of ice sheets and ice shelves is typically represented by a Maxwell visco-elastic model with nonlinear stress--dependent viscosity referred to as Glen's flow law \citep{glen_creep_1955,cuffey_physics_2010}. In the Maxwell model it is assumed that the total strain in a material body $\Omega_B$ can be decomposed additively into the elastic and viscous strains, whereas the total Cauchy stress acting on $\Omega_B$ is equal to both the elastic and viscous stresses:
\begin{align*}
\bm{\varepsilon} &= \bm{\varepsilon}_e+\bm{\varepsilon}_v,\\
\bm{\bm{\sigma}} &= \bm{\bm{\sigma}}_e = \bm{\bm{\sigma}}_v.
\end{align*}

The elastic stress $\bm{\bm{\sigma}}_e$ is typically \citep{greve_dynamics_2009} related to the elastic strain $\bm{\varepsilon}_e$ through Hooke's law
\begin{align}
\bm{\bm{\sigma}}_e = \lambda \tr(\bm{\varepsilon}_e) \mathds{1} + 2\mu \bm{\varepsilon}_e. 
\end{align}
Here $\mathds{1}$ is the second order identity tensor and  $\lambda, \mu$ are the first and second Lam\'e parameters that are related to the Young's modulus $E$ and Poisson ratio $\nu$:
\begin{align*}
\lambda = \frac{E \nu}{(1+\nu)(1-2\nu)}, \quad & \mu = \frac{E}{2(1+\nu)}.
\end{align*}
Alternatively, we may introduce the effective pressure $p$ that is, in the compressible case, related to the volumetric component of the elastic stress $p= -\left(\lambda + \frac{2\mu}{3} \right) \tr(\bm{\varepsilon}_e)$. Then the elastic stress is given by
\begin{align}\label{eq:elastic_stress_dual}
\bm{\bm{\sigma}}_e = -p \mathds{1} + 2\mu \dev(\bm{\varepsilon}_e), 
\end{align} 
where $\dev(\bm{\varepsilon}_e) = \bm{\varepsilon}_e - \frac{1}{3} \tr(\bm{\varepsilon}_e) \mathds{1}$ is the deviatoric part of the elastic strain tensor. The volumetric part is the effective pressure $p$ and deviatoric part $\dev(\bm{\sigma}_v)$ is governed by the Glen's flow law \citep{glen_creep_1955},
\begin{align}
\bm{\bm{\sigma}}_v = -p \mathds{1} + \dev(\bm{\sigma}_v) = -p \mathds{1} + 2 \eta(\dot{\bm{\varepsilon}}_v) \dot{\bm{\varepsilon}}_v.
\end{align}
The effective pressure can be expressed as $p = -1/3 \tr(\bm{\bm{\sigma}}_v)$. The viscous strain rate $\dot{\bm{\varepsilon}}_v$ is typically given by $\nabla_s \dot{\mathbf{w}}$, the symmetric gradient  of $\dot{\mathbf{w}}$, where $\dot{\mathbf{w}}$ is the rate of change of $\mathbf{w}$, and $\mathbf{w}$ is the irrecoverable part of the total displacement $\mathbf{u}$. The viscous stress $\bm{\bm{\sigma}}_v$ is decomposed into the volumetric and deviatoric parts. In effect, Glen's flow law \citep{glen_creep_1955} defines a viscosity that depends upon strain rate as follows.
\begin{align}
\eta(\dot{\bm{\varepsilon}}_v) = \frac{1}{2} {A}^{-1/n} \left(\frac{1}{2} \dot{\bm{\varepsilon}}_v : \dot{\bm{\varepsilon}}_v   \right)^{\frac{1-n}{2n}},
\end{align}
with $n$ usually considered to be $n=3$ and ${A}$ is a constant dependent on temperature via the Arrhenius law \citep{cuffey_physics_2010}. Note that the viscosity $\eta$ can be equivalently represented as a function $\widetilde{\eta}$ with dependency on the viscous shear stress \citep{greve_dynamics_2009},
\begin{align}\label{eq:viscosity_stress}
\eta(\dot{\bm{\varepsilon}}_v) = \widetilde{\eta}(\dev(\bm{\sigma}_v)) = \frac{1}{2} {A}^{-1} \left(\frac{1}{2} \dev(\bm{\sigma}_v) : \dev(\bm{\sigma}_v)   \right)^{\frac{1-n}{2}}.
\end{align}

Note that in our notation $\dot{\bm{\varepsilon}}_v$ is deviatoric due to the Glen's flow law. In certain places it may prove beneficial to write $\dev(\dot{\bm{\varepsilon}}_v)$ to highlight such a fact.

It proves useful to define the free energy and dissipation potential functions of the visco-elastic system.
The free energy function of the system is the recoverable elastic energy:
\begin{align}\label{eq:energy_storage_0}
\begin{split}
\psi(\bm{\varepsilon}_e) &= \frac{1}{2} \left( \lambda \tr(\bm{\varepsilon}_e)^2 + 2\mu \ \bm{\varepsilon}_e : \bm{\varepsilon}_e \right) \\
&= \frac{1}{2}\left(\left(\lambda +\frac{2\mu}{3}\right) \tr(\bm{\varepsilon}_e)^2 + 2\mu \dev(\bm{\varepsilon}_e) : \dev(\bm{\varepsilon}_e) \right).
\end{split}
\end{align}
We note that the elastic stress is given by $\bm{\bm{\sigma}}_e = \partial_{\bm{\varepsilon}_e} \psi(\bm{\varepsilon}_e).$

The dissipation potential {function} is given by:
\begin{align}
\phi(\dot{\bm{\varepsilon}}_v) = {\frac{2 n}{n+1}} \eta(\dev(\dot{\bm{\varepsilon}}_v)) \dev(\dot{\bm{\varepsilon}}_v) : \dev(\dot{\bm{\varepsilon}}_v).
\end{align}
The reason for defining the free energy and dissipation of the visco-elastic system is many fold. Primarily, it allows us to incorporate the fracture variable into the system and easily alter the system for different rheological assumptions. Secondly, it allows us to understand thermodynamic consistency of the system \citep{miehe_thermodynamically_2010}. Finally, it serves as a basis for the numerical implementation.\\

The stored energy and dissipation potential functionals of the system are the integral of the free energy and dissipation potential functions over the domain $\Omega_B$ \citep{maugin_internal_1990}:
\begin{align}
\mathbf{E} &= \int_{\Omega_B} \psi \diff V,\\
\mathbf{D} &= \int_{\Omega_B} \phi \diff V.
\end{align}
The free energy stored in $\Omega_B$ due to the deformation is given by $\mathbf{E}$. $\mathbf{D}$ is related to the power dissipated within the material during its deformation. \\

We assume that the macroscopic motions of the body $\Omega_B$ are given by the displacement field $\mathbf{u} \in \mathbb{R}^3$. In the small strain context we assume that the total strain is given by the symmetric part of the displacement gradient
\begin{align*}
\bm{\varepsilon} = \nabla_s \mathbf{u}.
\end{align*}
We further assume that the total displacement $\mathbf{u}$ can be additively decomposed into the elastic and viscous parts
\begin{align*}
\mathbf{u} = \mathbf{v} + \mathbf{w}.
\end{align*}
Therefore,
\begin{align*}
\bm{\varepsilon} = \nabla_s \mathbf{u} = \nabla_s \mathbf{v} + \nabla_s \mathbf{w} = \bm{\varepsilon}_e +\bm{\varepsilon}_v.
\end{align*}
In order to derive the system of equations governing the behaviour of the viscoelastic material, we choose the independent constitutive state variables to be $\mathbf{u}, \mathbf{w}$. We will make a standard assumption that the external forces act only on the total displacements \citep{miehe_multi-field_2011}. \\

We could also have chosen the total and elastic displacements ($\mathbf{u}$ and $\mathbf{v}$) as state variables (instead of $\mathbf{u}$ and $\mathbf{w}$). This would have lead to a Stokes-like system of equations.\\

The constitutive functions $\mathbf{E}$ and $\mathbf{D}$, defined above, are related respectively to energy storage and dissipation due to the deformation of the material \citep{miehe_multi-field_2011}. The rate of energy storage at state $\mathbf{u}, \mathbf{w}$ is the time derivative of the energy functional.
\begin{align*}
\frac{\diff}{\diff t} \mathbf{E} = \int_{\Omega_B} \delta_{\mathbf{u}} \psi(\mathbf{u},\mathbf{w}) \cdot \dot{\mathbf{u}} + \delta_{\mathbf{w}} \psi(\mathbf{u},\mathbf{w}) \cdot \dot{\mathbf{w}} \diff V + \int_{\partial_N \Omega_B} \left[\partial_{\nabla_s \mathbf{u}} \psi \cdot \mathbf{n}\right] \cdot \dot{\mathbf{u}} \diff S,
\end{align*}
where we introduced the functional derivatives of the free-energy function
\begin{align*}
\delta_{\mathbf{u}} \psi := - \nabla \cdot \left[\partial_{\nabla_s \mathbf{u}} \psi\right], \quad \delta_{\mathbf{w}} \psi := - \nabla \cdot \left[\partial_{\nabla_s \mathbf{w}} \psi\right].
\end{align*}
Note that the functional derivatives do not contain terms with $\partial_{\mathbf{u}}, \ \partial_{\mathbf{w}}$ due to requirements of frame invariance \citep{maugin_internal_1990,miehe_multi-field_2011}.\\

{The internal potential} $\Pi_{\mathrm{int}}$ is then composed of the elastically-stored and dissipated contributions as follows,
\begin{align} 
\Pi_{\mathrm{int}} &= \frac{\diff}{\diff t} \mathbf{E} + \mathbf{D}.
\end{align}
As such, the {internal potential} is determined by both the energy storage function $\psi$ and the dissipation potential function $\phi$. The effect is to combine the influence of the total displacement $\mathbf{u}$ and the viscous displacement $\mathbf{w}$.\\

We further assume that the {external load functional} is given by body and surface forces $\mathbf{f},\ \mathbf{t}$ that act only on the external variable $\mathbf{u}$
\begin{align} \label{eq:external_power}
\Pi_{\mathrm{ext}} &= \int_{\Omega_B} \mathbf{f} \cdot \dot{\mathbf{u}} \diff V + \int_{\partial_N \Omega_B} \mathbf{t} \cdot \dot{\mathbf{u}} \diff S.
\end{align}

We define the potential $\Pi$ as the difference between the internal potential functional and external load functional:
\begin{align}\label{eq:virtual_power}
\begin{split}
\Pi =  \Pi_{\mathrm{int}} - \Pi_{\mathrm{ext}} = &\int_{\Omega_B} \left[\delta_{\mathbf{u}} \psi - \mathbf{f}\right] \cdot \dot{\mathbf{u}} + \delta_{\mathbf{w}} \psi \cdot \dot{\mathbf{w}} +\phi \diff V \\ &+ \int_{\partial_N\Omega_B} \left[\partial_{\nabla_s \mathbf{u}} \psi \cdot \mathbf{n} - \mathbf{t} \right] \cdot \dot{\mathbf{u}} \diff S
\\ &+ \int_{\partial_N \Omega_{B}} \left[\partial_{\nabla_s \mathbf{w}} \psi \cdot \mathbf{n} \right] \cdot \dot{\mathbf{w}} \diff S.
\end{split}
\end{align}
On thermodynamic grounds and related principles \citep{miehe_multi-field_2011,maugin_internal_1990} we assume that the rates of the external and internal variables at a given state are determined by the variational principle
\begin{align}\label{eq:var_prin}
\lbrace \dot{\mathbf{u}},\ \dot{\mathbf{w}} \rbrace = \mathrm{Arg} \lbrace \inf_{\dot{\mathbf{u}},\ \dot{\mathbf{w}}} \Pi(\dot{\mathbf{u}},\ \dot{\mathbf{w}})  \rbrace.
\end{align}
Taking the variation of $\Pi$ we obtain an expression for virtual rates of the internal and external variables which satisfy homogeneous Dirichlet boundary conditions on $\partial_D \Omega_B$: 
\begin{align*}
\dot{\mathbf{u}} \in \lbrace \mathbf{v} | \mathbf{v} = \mathbf{0} \ \mathrm{on} \ \partial_D \Omega_B \rbrace, \quad \dot{\mathbf{w}} \in \lbrace \mathbf{v} | \mathbf{v} = \mathbf{0} \ \mathrm{on} \ \partial_D \Omega_B \rbrace.
\end{align*}

Applying the fundamental lemma of the calculus of variations then results in a coupled system of equations in a domain $\Omega_B$ with Neumann type boundary conditions:
\begin{equation}
\begin{aligned}
- \nabla \cdot \left( \lambda \nabla \cdot \left(\mathbf{u-w}\right)\mathds{1} +2 \mu \nabla_s \left(\mathbf{u-w}\right) \right) &= \mathbf{f} && \mathrm{in}\ \Omega_B \\
2\eta(\dev(\nabla_s \dot{\mathbf{w}}))\dev(\nabla_S \dot{\mathbf{w}}) - 2 \mu \dev \left( \nabla_s \left(\mathbf{u-w}\right) \right) &= 0 && \mathrm{in}\ \Omega_B \\
\nabla \cdot \dot{\mathbf{w}} &= 0 && \mathrm{in}\ \Omega_B \\
\left( \lambda \nabla \cdot \left(\mathbf{u-w}\right)\mathds{1} +2 \mu \nabla_s \left(\mathbf{u-w}\right) \right) \cdot \mathbf{n} &= \mathbf{t} && \mathrm{on} \ \partial_N \Omega_B.
\end{aligned}
\end{equation}
The first equation corresponds to the momentum balance equation of the elastic stress. The second equation corresponds to the balance of elastic and viscous deviatoric stresses. The third equation is the standard incompressibility condition and the fourth equation is a traction boundary condition.\\

The system can be also be rewritten using \eqref{eq:elastic_stress_dual} in the momentum balance equation
\begin{equation}
\begin{aligned}
- \nabla \cdot \left( -p \mathds{1} +2 \mu \dev\left(\nabla_s \left(\mathbf{u-w}\right) \right)\right) &= \mathbf{f} && \mathrm{in}\ \Omega_B \\
2\eta(\dev(\nabla_s \dot{\mathbf{w}}))\dev(\nabla_s \dot{\mathbf{w}}) - 2 \mu \dev \left( \nabla_s \left(\mathbf{u-w}\right) \right) &= 0 && \mathrm{in}\ \Omega_B \\
\nabla \cdot \left(\mathbf{u-w}\right) + \left(\lambda + \frac{2\mu}{3} \right)^{-1} p &=0 && \mathrm{in}\ \Omega_B\\
\nabla \cdot \dot{\mathbf{w}} &=0 && \mathrm{in}\ \Omega_B \\
\left( -p \mathds{1} +2 \mu \dev\left(\nabla_s \left(\mathbf{u-w}\right) \right)\right) \cdot \mathbf{n} &= \mathbf{t} && \mathrm{on} \ \partial_N \Omega_B.
\end{aligned}
\end{equation}

In the case of nearly incompressible materials, when the Poisson ratio $\nu \to 0.5$, $\lambda$ is much larger than $\mu$. This leads to a well known volume locking phenomenon \citep{babuska_locking_1992}.
This can be avoided by introduction of the pressure variable \citep{braess_finite_2007}.\\
To ensure stability of the solution, the inf-sup stability condition needs to be fulfilled \citep{braess_finite_2007}. This means that the discretisation of the viscoelastic system needs to be carefully chosen, see \citep{braess_finite_2007}.
\\

We assume that the external loading functions $\mathbf{f},\ \mathbf{t}$ are the gravitational force and depth varying hydrostatic water pressure, respectively, given by
\begin{align}\label{eq:boundary_conditions}
\mathbf{f} &= \rho_s \mathbf{g}, \ \mathrm{in} \ \Omega_B\\
\mathbf{t} &= - p_w \mathbf{n}, \ \mathrm{on} \ \partial_E \Omega_B \cup \partial_N \Omega_B
\end{align}
where $\rho_s$ is ice density, $\mathbf{g}$ is gravitational acceleration, $\mathbf{n}$ is the unit outward pointing normal vector, and $p_w$ is depth varying water pressure
\begin{align}\label{eq:water_pressure}
p_w = \begin{cases} \rho_w |\mathbf{g}| (H_w - z), \ &\mathrm{for}\ z< H_w \\
0, &\mathrm{for} \ z \geq H_w
\end{cases}
\end{align}
with $H_w = \rho_s/\rho_w H$ and $H$ being the ice-shelf thickness. Throughout this section we focused on an intact material body without any inclusion or cracks. Therefore, the traction boundary condition is acting only on the exterior boundaries as no cracks exist inside of the material $\Omega_B$. Water filled cracks will be incorporated into the system via the phase field variable in Subsection~\ref{ssec:phase_field}.

\subsection{Phase field fracture formulation}\label{ssec:phase_field} 

Before describing the details of the implementation we first highlight the principal differences from the situation without fracture considered in the previous section. We will briefly introduce the energetic approach to fracture, then extend the equations from Subsection~\ref{ssec:ice_rheology} to incorporate fracture into the viscoelastic constitutive equations via a phase-field variable $d$ that takes values of $d=1$ near the cracks and $d=0$ away from the cracks. In doing so, we eliminate treatment of evolving geometry due to fracture evolution. Instead, our equations will be defined over the whole extended domain $\Omega$ instead of $\Omega_B$.\\

In the standard theory of brittle fracture, the drop in stored elastic energy $G$ that occurs when a specific crack $\mathcal{C}$ is introduced into the material, is compared to the critical energy $G_c$ that is needed to create the crack. Propagation of the crack $\mathcal{C}$ occurs when $G \geq G_c$. This is known as the Griffith criterion \citep{griffith_vi_nodate}. For pure brittle failure $G_c$ is envisioned to be the surface energy that is required to separate the crack faces, but this can also be generalised for more ductile materials, in which additional energy must be expended to form the crack. The resulting material parameter $G_c$ is commonly known as the Griffith energy release rate and can be estimated from laboratory experiments. \\

In the phase-field approach, outlined in \citep{francfort_revisiting_1998,miehe_phase_2010,miehe_multi-field_2011}, the energy $\Gamma(\mathcal{C})$ needed to create a crack $\mathcal{C}$ is approximated by an elliptic functional that depends upon the critical energy release rate $G_c$, the phase-field variable $d$, and its spatial gradient $\nabla d$,
\begin{align}
\Gamma_{\ell}(d) = \int_{\Omega} G_c \gamma(d,\nabla d) \diff V.
\end{align} 
The crack density function $\gamma$ approximates the specific surface area of cracks per unit volume and is defined by
\begin{align}
\gamma(d, \nabla d) = \frac{1}{2\ell} \left( d^2 + \ell^2 |\nabla d|^2 \right).
\end{align}
 In practice, the regularisation parameter $\ell$ acts to control the lengthscale over which the phase-field variable $d$ varies in the neighborhood of cracks.  Larger $\ell$ corresponds to a smoother regularised transition between fully-fractured material ($d=1$) and fully-intact material ($d=0$). It is this regularisation that allows sharp cracks to be represented on a finite computational mesh. Although approximate, there are theoretical reasons to expect $\Gamma_{\ell}$ to become ever closer to the true energy $\Gamma$ as $\ell$ is decreased \citep{ambrosio_approximation_1990}.\\

To capture the release of elastic energy upon crack formation, alternative degraded forms for the free-energy function $\widetilde{\psi}(\bm{\varepsilon}_e,d)$ and the dissipative potential $\widetilde{\phi}(\dot{\bm{\varepsilon}}_v,d)$ are used. These functions play the same role as $\psi(\bm{\varepsilon}_e)$ and $\phi(\dot{\bm{\varepsilon}}_v)$ in the previous section, but now account for reduced ability to maintain elastic or viscous stresses within the cracked material. \\

The net result is that we can define {the internal potential} of the system as modified by the presence of the phase field $d$ as follows:

\begin{align}
\begin{split}
\Pi_{\mathrm{int}} &= \frac{\diff}{\diff t} \widetilde{\mathbf{E}} +\widetilde{\mathbf{D}} +\frac{\diff}{\diff t}\Gamma_{\ell}\\
&= \frac{\diff}{\diff t} \int_{\Omega} \widetilde{\psi}(\bm{\varepsilon}_e,d) \diff V + \int_{\Omega} \widetilde{\phi}(\dot{\bm{\varepsilon}}_v,d) \diff V +\frac{\diff}{\diff t} \int_{\Omega} G_c \gamma(d,\nabla d) \diff V,
\end{split}
\end{align}
where $\widetilde{\psi}(\bm{\varepsilon}_e,d)$ and $\widetilde{\phi}(\dot{\bm{\varepsilon}}_v,d)$ are the modified free-energy density and dissipation potentials, and $\gamma(d, \nabla d)$ is the crack density function. \\

In a similar fashion, we modify the {external load functional}, which we assume to be given as in \eqref{eq:external_power}, but with an extra term that represents the effects of pressurised water within the cracks.
\begin{align} \label{eq:internal_power_fracture}
\Pi_{\mathrm{ext}} = \int_{\Omega} {g(d)}\mathbf{f} \cdot \dot{\mathbf{u}} \diff V + \int_{\partial_N \Omega} {g(d)} \mathbf{t}\cdot \dot{\mathbf{u}} \diff S \ {+} \int_{\Omega} p_w {\nabla g(d)} \cdot \dot{\mathbf{u}} \diff V
\end{align}

In this expression, external forces have been multiplied by a degradation function $g(d)$ that takes a value $g=1$ in fully-intact regions and $g=0$ in fully-fractured regions. The intention is to approximate \eqref{eq:external_power}, in which the external forces do not contribute within the cracked material. In line with the literature \citep{miehe_phase_2010} we choose 
\begin{align}\label{eq:degradation_fn}
g(d) = (1-d)^2.
\end{align}
This is dependent on the phase-field variable $d$ and smoothly interpolates between the undamaged state $d = 0$ and a fully damaged state $d=1$. The motivation for this choice of function is that it satisfies the following criteria,
\begin{align*}
g(0) = 1, \quad g(1) = 0, \quad g'(1) = 0.
\end{align*}
These conditions impose limits on the amount of degradation for the undamaged and fully damaged states. The last condition ensures that the fracture force converges to a finite value when $d \to 1$ \citep{miehe_phase_2010}. \\

The final term in \eqref{eq:internal_power_fracture} is motivated by an approximation that allows the effects of water pressure on internal crack faces to be represented as a volume integral
\begin{align}
{\int_{\cup_i \partial \mathcal{C}_i} p_w \mathbf{n} \cdot \dot{\mathbf{u}} \diff S \approx  - \int_{\Omega} p_w \nabla g(d) \cdot \dot{\mathbf{u}} \diff V = 
\int_{\Omega} 2 p_w (1-d) \nabla d \cdot \dot{\mathbf{u}} \diff V.}
\end{align}
This approximation is derived in Appendix~\ref{app:phase_field}.\\

Turning to the choice of $\widetilde{\psi}(\bm{\varepsilon}_e,d)$, a straightforward degradation of free energy $\psi(\bm{\varepsilon}_e)$ could be obtained simply by multiplying this functions by $g(d)$. However, this would lead to unrealistic behaviour due to equal treatment of fracture under tension and compression. In order to alleviate this issue we seek a tensile--compressive decomposition so that the intact material behaves in line with equations in Subsection~\ref{ssec:ice_rheology}, but in fully damaged parts of the material should hold no or almost no tensile stresses.\\

Inspired by \citep{miehe_phase-field_2017,miehe_phase_2010} we split the energy storage function into tensile and compressive parts as follows:
\begin{align}
\widetilde{\psi} (\bm{\varepsilon}_e,d) = g(d) \left( \psi^+(\bm{\varepsilon}_e) - \psi_{crit} \right) + \left( \psi^-(\bm{\varepsilon}_e)+\psi_{crit}\right).
\end{align}
Here, $\psi(\bm{\varepsilon}_e) = \lambda/2 \tr(\bm{\varepsilon}_e)^2 + \mu \tr(\bm{\varepsilon}_e^2) $ corresponds to an isotropic energy function of an unbroken material and $\psi_{crit}$ is a material parameter that acts as a crack energy threshold \citep{miehe_phase-field_2017}. Note that the crack energy threshold has no impact on the intact material, where $d = 0$. In the case of a fully damaged material, $\psi_{crit}$ acts as a regularisation parameter that prevents a complete degradation of the tensile energy-density function.\\

We consider an additive decomposition of $\psi = \psi^+ + \psi^-$ where
\begin{align}
\psi^{\pm} &= \frac{1}{2}\left(\lambda+\frac{2\mu}{3}\right) \langle \tr(\bm{\varepsilon}_e)\rangle_{\pm}^2 +\mu \tr\left({\left(\dev(\bm{\varepsilon}_{e})_{\pm}\right)^2}\right),
%\psi^{\pm} &= \frac{1}{2}\left(\left(\lambda+2\mu/3\right)\langle \tr(\bm{\varepsilon}_e)\rangle_{\pm}^2 +2\mu  \tr\left({\dev(\bm{\varepsilon}_e)_{\pm}^2}\right) \right) \\
%&= \frac{1}{2} \left(  \left(\lambda+2\mu/3\right)^{-1}\langle -p \rangle_{\pm}^2 +2\mu  \tr\left({\dev(\bm{\varepsilon}_e)_{\pm}^2}\right) \right)
\end{align}
where $\langle \cdot \rangle_{\pm} $ is the Macaulay bracket given by $\langle \cdot \rangle_{\pm} = 1/2 (\cdot \pm |\cdot|)$ and $\bm{\varepsilon}_{\pm}$ is a spectral decomposition of $\bm{\varepsilon}$ into positive and negative parts. The spectral decomposition is given by
\begin{align*}
\bm{\varepsilon}_{\pm} := \sum_{a} \langle e_a \rangle_{\pm} \mathbf{m}_a \otimes \mathbf{m}_a,
\end{align*}
where $e_a$ are the principal strains and $\mathbf{m}_a$ are the principal directions. \\

Next, we define the dissipative potential $\widetilde{\phi}(\dot{\bm{\varepsilon}}_v,d)$. Contrary to the free energy, we do not decompose the dissipative potential function into tensile and compressive parts, but simply degrade the dissipative potential $\phi(\dot{\bm{\varepsilon}}_v)$ as follows,
\begin{align}
\widetilde{\phi}(\dot{\bm{\varepsilon}}_v,d) = g(d) \phi(\dot{\bm{\varepsilon}}_v) = g(d) {\frac{2 n}{n+1}} \eta(\dev(\dot{\bm{\varepsilon}}_v)) \dev(\dot{\bm{\varepsilon}}_v) : \dev(\dot{\bm{\varepsilon}}_v).
\end{align}

Equipped with the modified expressions for $\Pi_{\mathrm{int}}$ and $\Pi_{\mathrm{ext}}$, we proceed as in Subsection~\ref{ssec:ice_rheology}, using the same variational principle \citep{maugin_method_1980,maugin_internal_1990} to derive the strong form of the equations for the viscoelastic phase-field fracture system. Now, as well as all admissible virtual rates $\mathbf{\dot{u} , \ \dot{w}}$ we must also consider $\Pi_{\mathrm{int}}$ and $\Pi_{\mathrm{ext}}$ to balance for all admissible variations with respect to $\dot{d}$, the rate of change of the phase-field variable. This provides an additional equation that must be satisfied by the phase-field whenever $\dot{d}>0$,

\begin{equation}
\begin{aligned}
\frac{G_c}{\ell} \left( d-\ell^2 \Delta d\right) &= 2(1-d) \left( \psi^+ - \psi_{crit} \right) && \mathrm{in} \ \Omega, \\
\nabla d \cdot \mathbf{n} &= 0 && \mathrm{on} \ \partial \Omega.
\end{aligned}
\end{equation}

As in most models, fracture is designed to be irreversible, with a requirement on thermodynamic consistency \citep{miehe_thermodynamically_2010}. The rate of change of $\Gamma_{\ell}(d)$ defines the dissipation of power used in creating the crack field
\begin{align*}
\dot{\Gamma}_{\ell}(d) = \int_{\Omega} G_c \delta_d \gamma(d,\nabla d) \dot{d} \diff V,
\end{align*}
where $\delta_d \gamma(d,\nabla d)$ is the variational derivative  of the crack density function. \\

To ensure  thermodynamically-consistent, irreversible fracture, we require that the dissipated power is positive 
\begin{align*}
\dot{\Gamma}_{\ell}(d) \geq 0.
\end{align*}
This condition can be satisfied by locally imposing that 
\begin{align}
\delta_d \gamma(d,\nabla d) &\geq 0,\label{eq:irrev1}\\
\dot{d} &\geq 0.\label{eq:irrev2}
\end{align}
The second inequality is a natural constraint that assumes the phase-field evolution is locally irreversible and does not account for any healing. \\

 In order to satisfy the conditions \eqref{eq:irrev1} and \eqref{eq:irrev2}, we follow \citep{miehe_thermodynamically_2010} and introduce a local history field of maximum tensile energy over time or loading steps
\begin{align}\label{eq:history_fn}
\mathcal{H}(x,t) := \max_{s} \langle \psi^+(\bm{\varepsilon}_e(x,s))-\psi_{crit}\rangle_+ .
\end{align}
The fracture phase-field evolution is then modified to be

\begin{equation}
\begin{aligned}
\frac{G_c}{\ell} \left( d-\ell^2 \Delta d\right) &= 2(1-d) \mathcal{H} && \mathrm{in} \ \Omega, \\
\nabla d \cdot \mathbf{n} &= 0 && \mathrm{on} \ \partial \Omega.
\end{aligned}
\end{equation}
We will use this formulation for the implementation. \\

For completeness, we present the full system of equations derived from the modified expressions for $\Pi_{\mathrm{int}}$ and $\Pi_{\mathrm{ext}}$ below. To simplify the dependence of these equations upon material parameters, we present them in non-dimensional form. This is obtained by introducing non-dimensional scaled quantities (labelled by asterix) defined as follows:
\begin{align}
\begin{split}
&\mathbf{x} = L \mathbf{x}^*,
\quad t = \tau t^*,
\quad \mathbf{u} = u_c \mathbf{u}^*,
\quad \mathbf{w} = u_c \mathbf{w}^*,
\quad \ell = L \ell^*, \\
&\lambda = \mu_c \lambda^*, 
\quad \mu = \mu_c \mu^*,
\quad p_w = p_{wc} p_w^*,
\quad \mathbf{f} = f_c \mathbf{f}^*,
\quad \mathbf{t} = p_c \mathbf{t}^*, \\
&\rho_s = \rho_c \rho_s^*,
\quad \rho_w = \rho_c \rho_w^*,
\quad \eta = \eta_c \eta^*, 
\quad \bm{\bm{\bm{\sigma}}} = p_c \bm{\bm{\sigma}}^* , 
\quad p = p_c p^*,\\
&\mathcal{H} = \mathcal{H}_{c}\mathcal{H}^*,
\quad \psi = \mathcal{H}_{c} \psi^*, 
\quad \psi^{\pm} = \mathcal{H}_{c} \psi^{\pm *}, 
\quad \psi_{crit} = \mathcal{H}_{c} \psi_{crit}^*.
\end{split}
\end{align}
The scaling of $x,\ y,\ z$ is to be understood as the scaling of all physical dimensions by a characteristic scale for the ice thickness $L$. Similarly, time is scaled by $\tau$ and displacements by $u_c$. The other scales are chosen in terms of $L$, $\tau$, $u_c$ and material constants as follows: 

\begin{align*}
p_c = \mu u_c L^{-1}, 
\, \eta_c &= A^{-\frac{1}{n}} \left( \frac{u_c}{L\tau}\right)^{\frac{1-n}{n}}, \, \mathcal{H}_{c} = \mu \left(\frac{u_c}{L}\right)^2,
\, \mu_c = \mu,
\, f_c = \rho_c g,
\, p_{wc} = \rho_c g L.
\end{align*}

Having derived the full system in exactly the same way as described previously for the situation without fracture, we substitute these relationships, then drop the asterix on non-dimensional quantities to provide the following non-dimensional system of equations.
\begin{equation}\label{eq:full_system_nondim}
\begin{aligned}
- \nabla \cdot  \bm{\bm{\sigma}} &=  C_1 \left[ g(d) \mathbf{f} - p_w \nabla g(d)\right] && \mathrm{in}\ \Omega \\
g(d) 2\eta(\dev(\nabla_s \dot{\mathbf{w}}))\dev(\nabla_S \dot{\mathbf{w}}) - 2 C_2  \dev(\bm{\bm{\sigma}}) &= 0 && \mathrm{in}\ \Omega \\
\nabla \cdot \left(\mathbf{u-w}\right) + \frac{3(1-2\nu)}{2(1+\nu)} p &=0 && \mathrm{in}\ \Omega\\
\nabla \cdot \dot{\mathbf{w}} &=0 && \mathrm{in}\ \Omega \\
\bm{\bm{\sigma}} \cdot \mathbf{n} &= g(d) \mathbf{t} && \mathrm{on} \ \partial_N \Omega\\
 d-\ell^{2} \Delta d &= C_3 \ell 2(1-d) \mathcal{H} && \mathrm{in} \ \Omega \\
\nabla d \cdot \mathbf{n} &= 0 && \mathrm{on} \ \partial \Omega.
\end{aligned}
\end{equation}
where we have defined $\bm{\bm{\sigma}} = g(d)\bm{\bm{\sigma}}_e^+ + \bm{\bm{\sigma}}_e^-$ with $\bm{\bm{\sigma}}_e^{\pm} :=  \langle -p \rangle_{\pm} \mathds{1} +2 \dev\left(\nabla_s \left(\mathbf{u-w}\right)\right)_{\pm}$ for notational convenience. The history function $\mathcal{H}$ is defined as in \eqref{eq:history_fn}.

The only non-dimensional parameters that enter into the system are the Poisson ratio $\nu$, the regularisation length $\ell^*$, the threshold $\psi_{crit}^*$ that enters via \eqref{eq:history_fn}, and the constants $C_1, C_2, C_3$. These constants depend on the scales $L$, $\tau$, $u_c$, and other material and physical parameters as follows:
\begin{align*}
C_1 &= \frac{L^2\rho_c |\mathbf{g}|}{u_c \mu},\\
C_2 &= A^{\frac{1}{n}} \left(\frac{u_c}{L}\right)^{1-1/n}\mu \tau^{\frac{1}{n}},\\
C_3 &= \frac{\mu u_c^2}{G_c L}.
\end{align*}
These non-dimensional constants govern the operating regime of our equations. Constant $C_1$ corresponds to the ratio between external and elastic stresses, constant $C_2$ is the ratio between elastic and viscous stresses, and constant $C_3$ is the ratio between elastic stress and fracture stress. For increasing $C_1$, gravity and water pressure effects become more important to the problem. Increasing $C_2$ takes the model from fully elastic through visco-elastic to fully viscous. Increasing $C_3$ increases the role of fracture from being irrelevant towards a zero-stress failure criterion. This gives a sense of how changes in material parameters taken from the literature (Table~\ref{tab:parameters}) will influence the model. As an example, for fixed $C_2$, doubling the rate factor $A$ halves the characteristic timescale $\tau$.

For our simulations we have chosen the characteristic length-scale $L = 100m$, characteristic displacement $u_c = 10^{-2} m$, and characteristic timescale $\tau = 1 a$. This leads to nondimensional constants $C_1 = 2.79, \ C_2 = 11.82, \ C_3 = 3520.75$. The nondimesional crack regularisation length-scale is chosen as $\ell^* = 5\times 10^{-3}$ which corresponds to a dimensional quantity of $\ell = 0.5 m$. 

An alternative choice of scaling, not considered further here, would be to fix $L$ then choose scales $u_c$, and $\tau$ such that $C_1 = C_2 = 1$, and finally choose the non-dimensional regularisation length $\ell^*$ such that $ C_3 \ell^* = 1$. This would allow further simplification of the equations, leaving just three non-dimensional parameters; the Poisson ratio $\nu$, the regularisation length $\ell^*$, the threshold $\psi_{crit}^*$. The Poisson ratio for ice is known approximately \citep{greve_dynamics_2009}, while $\ell^*$, and $\psi_{crit}^*$ can be viewed as small parameters that regularise the system of equations. Because $\ell^{*2}$ multiplies the highest derivative of $d$, the problem has the character of a singular perturbation problem.

\begin{table}
\begin{center}
\begin{tabular}{|c|c|}
\hline
Parameters & Values\\
\hline
$\rho_s$ & $900 \ [kg/m^3]$\\
$\rho_w$ & $1000 \ [kg/m^3]$\\
$|\mathbf{g}|$ & $9.81 \ [m/s^2]$\\
$E$ & $9.33 \times 10^{9} \ [N/m^2]$ \\
$\nu$ & $0.325 \ [-]$ \\
$ A$ & $1.2 \times 10^{-25} \ [Pa^{-3} s^{-1}] $\\
$n$ & 3 [-]\\
$G_c$ & $1 \ [N/m]$\\
$\psi_{crit}$ & $1 \ [Pa]$\\
\hline 
\end{tabular}
\end{center}
\caption{Table of material parameters and their characteristic values. The parameters for Young's modulus $E$, and Poisson ratio $\nu$ are from \citep{greve_dynamics_2009}, flow constant $A$ is from \citep{cuffey_physics_2010}, and critical energy release rate $G_c$ is from \citep{goodman_critical_1980}.}\label{tab:parameters}
\end{table}

\section{Implementation Details}\label{sec:implementation}
In this section we briefly describe the implementation details of the viscoelastic phase field fracture system \eqref{eq:full_system_nondim}. The computational domain is decomposed into a triangular mesh and is refined in areas where cracks are expected to propagate. In order to obtain mesh independent results, the mesh size near the crack path needs to be chosen at least two times smaller than the regularisation parameter $\ell$. The weak formulation is implemented using a mixed finite element method in space, using piecewise polynomial functions to approximate the solution. In order to ensure numerical stability of the resulting system we use Taylor-Hood elements, where piecewise quadratic functions are used for the displacements $\mathbf{u},\ \mathbf{w}$ and piecewise linear functions for the pressure $p$ and phase field fracture variable $d$ \citep{mang_phase-field_2020}. In order to enforce the incompressibility condition  in \eqref{eq:full_system_nondim} we include a viscous pressure-like variable $q$ that will act as a Lagrange multiplier, and is approximated by piecewise linear functions. This way we also ensure symmetry of the finite element system. We approximate the time derivatives using implicit Euler time stepping method.\\

We note that the system \eqref{eq:full_system_nondim} is nonlinear and nonconvex due to the proposed split into tensile-compressive parts and due to the degradation function $g(d)$. Consequently, a discretised problem of \eqref{eq:full_system_nondim} would lead to a nonsymmetric linear system. In order to alleviate this problem we propose an extension of the staggered scheme \citep{miehe_phase_2010} on an augmented system of equations. We iteratively solve a system for the displacement-pressure, and the fracture phase field variable using a staggered alternating minimisation algorithm.\\

At each time step $k$ and at each step $i$ of the staggered iteration we consider the phase field variable $d$ fixed and solve the following system in order to obtain displacement and pressure variables:
\begin{align}\label{eq:lin_sys1}
\begin{bmatrix}
2C_2^{-1}/\Delta t\bm{\tilde{K}} +2\bm{K} & -2 \bm{K} & -\bm{V}^T & \bm{V}^T \\
-2 \bm{K} & 2 \bm{K} & \bm{0} & -\bm{V}^T \\
-\bm{V} & \bm{0} & \bm{0} & \bm{0} \\
\bm{V} & -\bm{V} & \bm{0} & -\frac{3(1-2\nu)}{2(1+\nu)} \bm{M} \\
\end{bmatrix}
\begin{bmatrix}
\mathfrak{w}^{(k,i)} \\
\mathfrak{u}^{(k,i)} \\
\mathfrak{q}^{(k,i)} \\
\mathfrak{p}^{(k,i)}
\end{bmatrix}
=
\begin{bmatrix}
2C_2^{-1}/\Delta t \bm{\tilde{K}}\mathfrak{w}^{(k-1,I)}\\
C_1\left( \bm{F} + \bm{P}\right) + \bm{T}  \\
-\bm{V}  \mathfrak{w}^{(k-1,I)} \\
\bm{0}
\end{bmatrix} 
\end{align}
The finite element matrices and vectors are defined in Appendix~\ref{app:numerics}. The time step is denoted $\Delta t$ and we take a uniform step of 1 day. The inner iterative process continues until the system converges to a steady state and the last iteration is denoted $I$.\\
We then treat the displacement and pressure variables as fixed and solve for the fracture phase field variable ${d}$:
\begin{align}\label{eq:lin_sys2}
(\bm{M}+2 C_3 \ell \hat{\bm{M}}-\ell^2 \hat{\bm{K}}) \mathfrak{d}^{(k,i)} = 2 C_3 \ell \bm{\mathcal{H}}^{(k,i)}.
\end{align}
The finite element matrices and vectors are again defined in Appendix~\ref{app:numerics}. The history function $\bm{\mathcal{H}}^{(k,i)}$ is computed from the solution of $\left( \mathfrak{w}^{(k,i)}, \mathfrak{u}^{(k,i)},\mathfrak{q}^{(k,i)},\mathfrak{p}^{(k,i)} \right)$ as defined in \eqref{eq:history_fn}.\\

Under the assumptions of small displacements and strains, linear rheology, and load independent boundary conditions, we can usually discount differences between the original and current configurations as the differences are negligibly small. However, for nonlinear rheological equations and load dependent boundary conditions as in \eqref{eq:full_system_nondim} extra care needs to be taken to obtain physically meaningful results, even under the assumptions of small displacements and small strains \citep{bathe_finite_1975,maugin_method_1980}. Furthermore, evolution over long time periods may invalidate the small displacement assumption. In such cases we need to address the differences between the body configurations at different times.\\
We adopt an approach described in \citep{bathe_finite_1975} referred to as material-nonlinearity-only based on the updated Lagrangian formulation, which allows us to distinguish the body configuration at different times, and is described in Appendix~\ref{app:numerics}.

\section{Numerical Experiments}\label{sec:numerics}
In this section we present numerical experiments to demonstrate the capabilities of the model. Our primary focus is on tabular icebergs that float freely in the water and are unconstrained laterally. We ignore any tidal and drift effects. \\

We consider a tabular iceberg with varied thickness $H = 100, 200, 300, 400, 500, 600 m$ and length of $16000m$. In each run we initialise the iceberg in a floating equilibrium with external forces switched off. We then turn on the external forces due to gravity and water pressure. The iceberg will initially deform elastically and then start to evolve due to viscous creep. 
We initialise the icebergs either without any notches present, with notches $5 m$ in depth and one ice thickness away from the ice front, or with notches uniformly spaced at $100 m$ intervals across the whole surface of the iceberg. Models are run for a period equivalent to $3$ years.

\begin{figure}
  \centering
\includegraphics[width=0.99\textwidth]{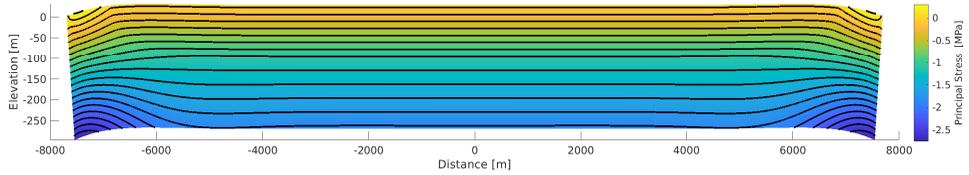}
\caption{The iceberg geometry at $t = 0$ days, showing the initial elastic displacement caused by interactions between gravity and water pressure. Displacements are exaggerated by a factor of 100.}\label{fig:initialdeformation}
\end{figure}

Figure~\ref{fig:initialdeformation} demonstrates the initial deformation and stress distribution for an iceberg with thickness $H = 300m$. Upon initiation of the model, there is a concentration of tensile stress at the ice surface due to bending with a maximum at a distance of just under one ice thickness from the ice front \citep{reeh_calving_1968}. Surface damage immediately appears at this stress maximum forming a crack with depth approximately equal to the Nye depth \citep{nye_comments_1955,weertman_can_1973}.

\begin{figure}
  \begin{minipage}[t]{1\linewidth}  
\subfloat[][$t=0$ days]{
\includegraphics[width=0.99\textwidth]{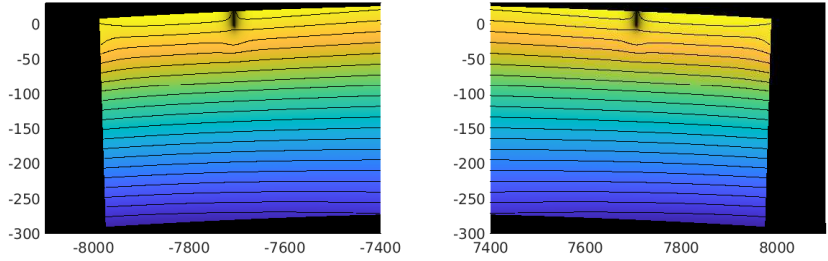}}\\
\subfloat[][$t=98$ days]{
\includegraphics[width=0.99\textwidth]{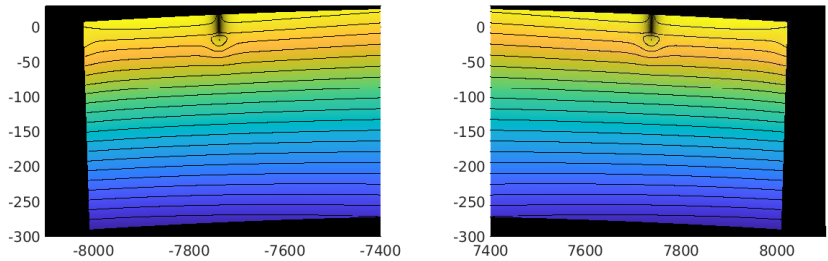}}\\
\subfloat[][$t=258$ days]{
\includegraphics[width=0.99\textwidth]{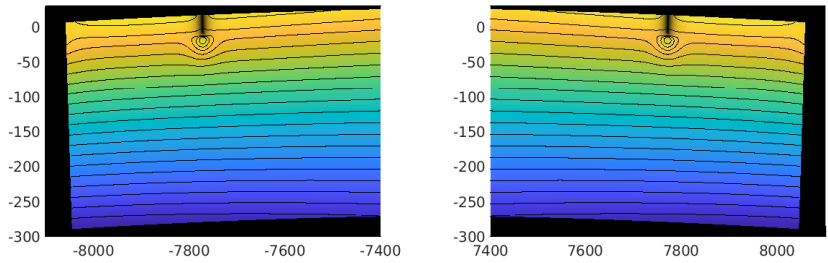}}\\
\subfloat[][$t=260$ days]{
\includegraphics[width=0.99\textwidth]{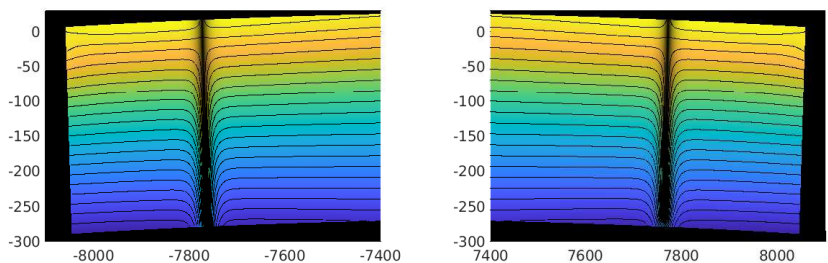}}\\
  \end{minipage}%
  \caption{Figure (a) details at $t=0$ days of iceberg geometry, principal stress (in colour) and phase field (shading to black at $d=1$ corresponding to fully fractured locations). Details are shown at both ends of the iceberg (displacements are exaggerated by a factor of 100). Figures (b) and (c) demonstrate the viscous creep of the iceberg and stress concentration near the crack tip at intermediate times of $t = 98$ and $ 258$ days, respectively. Figure (d) displays both ends of the iceberg at moment of calving ($t=260$ days).  }\label{fig:creep_deformation}
\end{figure}

This is a stable configuration that then evolves through creep, whereby the ice shelf spreads, thins and bends over time. Figure~\ref{fig:creep_deformation} demonstrates the temporal evolution of both ends of the iceberg. The initial crack length remains constant as the presence of the crack transfers stress to the remaining intact ice area below the crack. The stress concentration associated with the crack tip is clearly visible just below the base of the crevasse. During this phase of creep, the initial crack tip migrates towards the water line. After a finite time, the crack tip crosses the water line where the additional stress provided by water pressure provides a tensile component sufficient to cause rapid crack propagation. This ultimate stage of failure occurs within one time-step of the model and can be considered a type of hydro-fracture, occurring almost instantaneously.


Crack location and the timing of propagation both vary strongly with the ice thickness. The exact timing will also depend on the values of material constants (Table~\ref{tab:parameters}). For simulations with $n=3$ in the Glen flow law, the time from model initiation to calving is proportional to the ice thickness $H^{-3}$ while the distance from the crack to the ice front is proportional to $H$. This produces an effective calving rate that is proportional to $H^{4}$.

\begin{figure}
\centering
\subfloat[][]{
\includegraphics[width=0.45\textwidth]{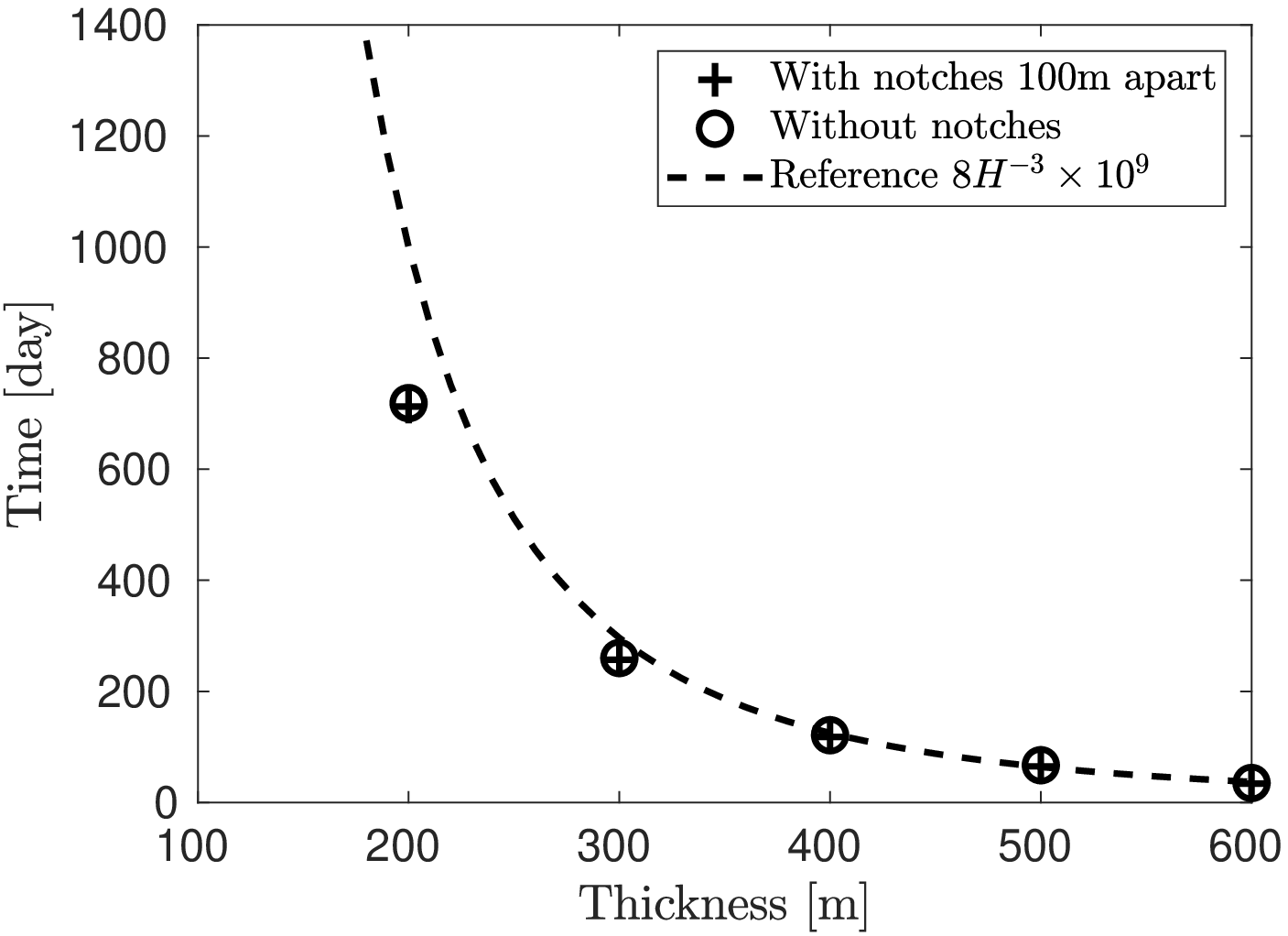}}
\subfloat[][]{
\includegraphics[width=0.45\textwidth]{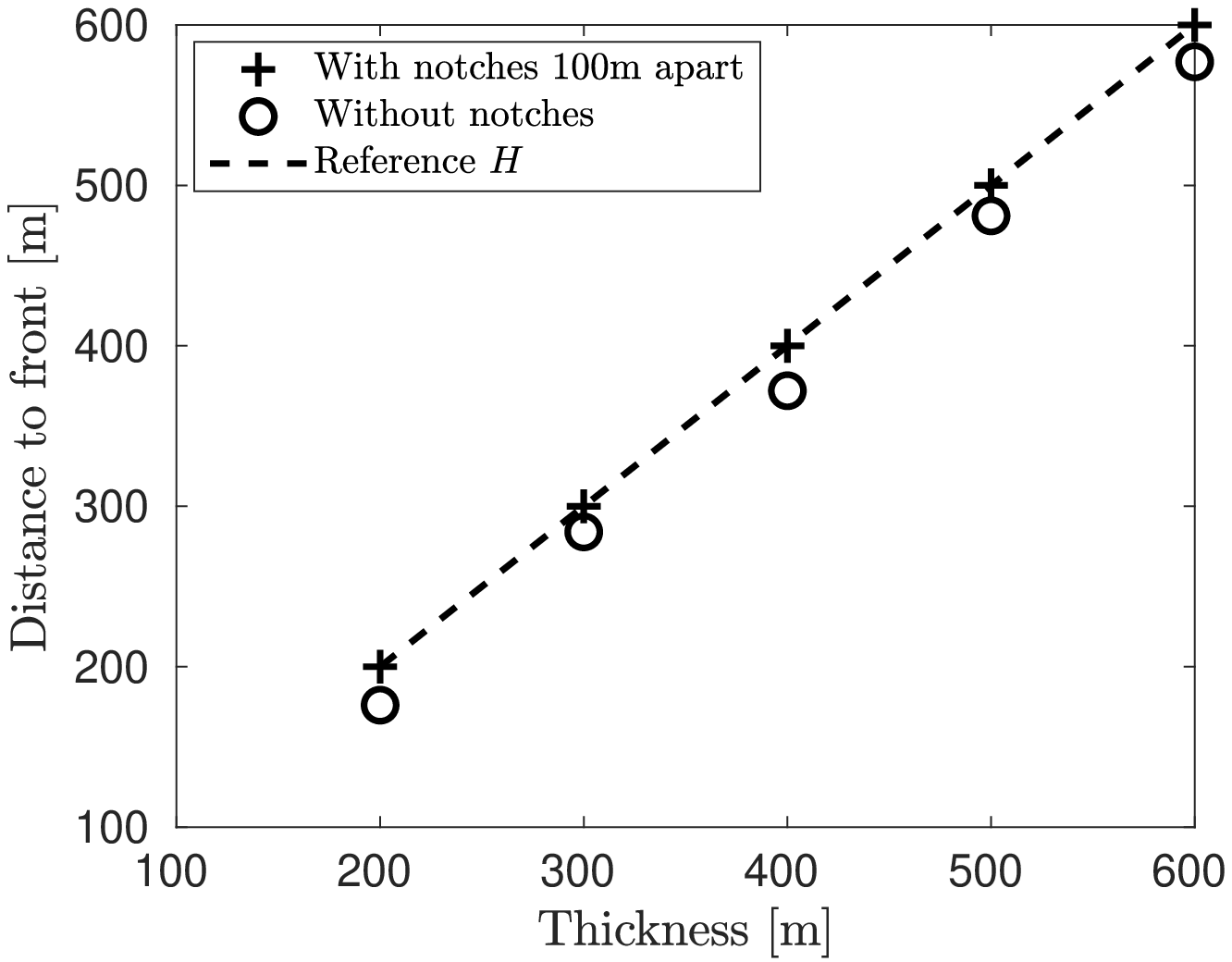}}\\
\subfloat[][]{
\includegraphics[width=0.45\textwidth]{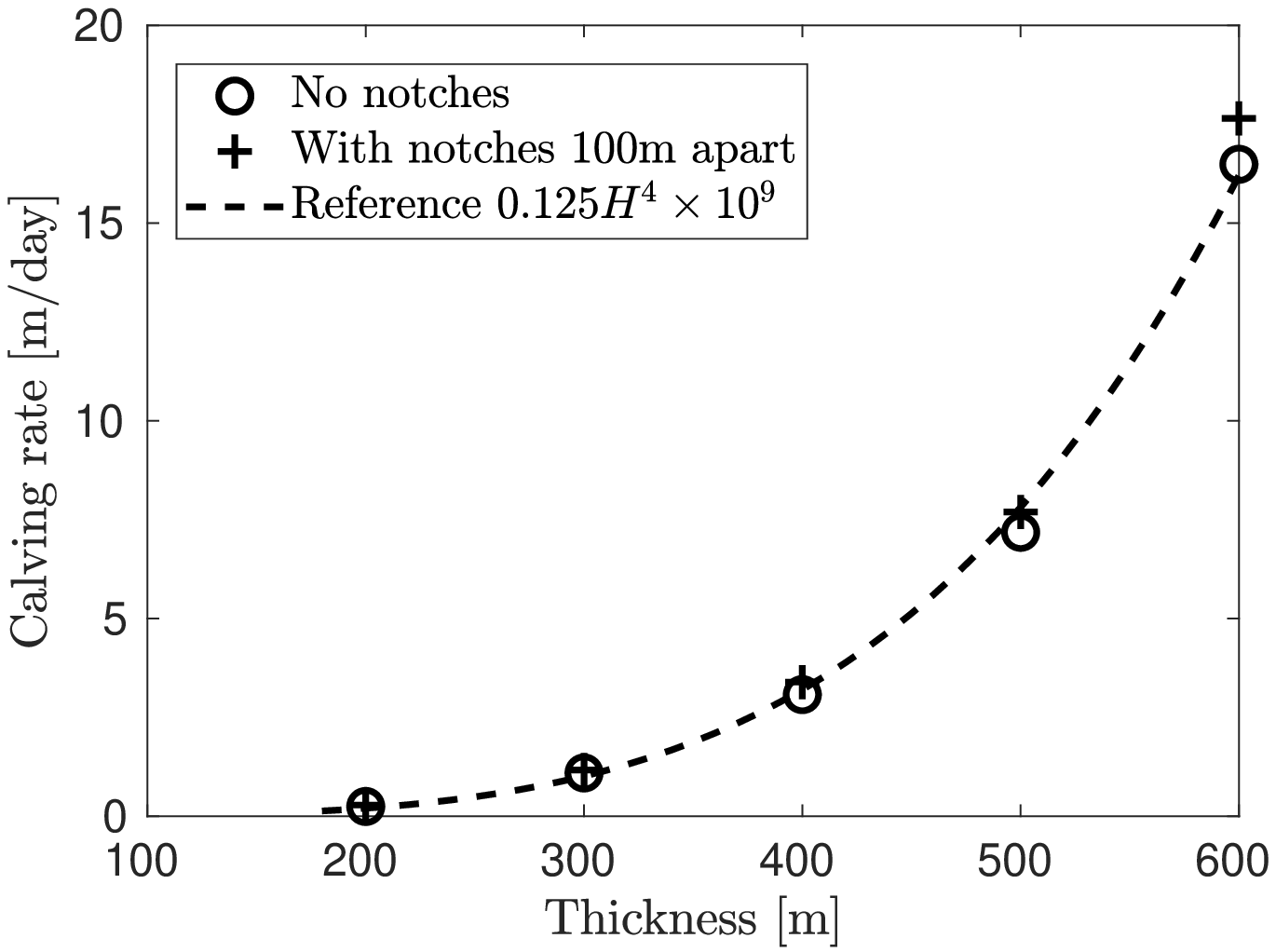}}
\caption{Results for icebergs of various initial thickness (a) time to first calving event; (b) horizontal dimension of first iceberg calved; (c) lateral calving rate from first calving. Results are shown with and without pre-imposed notches. Reference curves (dotted) show the predictions of a simple model in which strain thinning allows a fixed-depth crevasse to approach the water level, whereupon hydro-fracture occurs throughout the full depth. }\label{fig:timing}
\end{figure}

The inclusion of initial notches in the model domain has a negligible effect on the timing of calving (Figure~\ref{fig:timing}), although naturally initiated cracks occur slightly closer to the ice front than when notches are pre-defined at $100m$ intervals and icebergs with pre-defined notches calve slightly sooner than naturally initiated icebergs.

\section{Discussion}\label{sec:discussion}

The simulations demonstrate that even brittle failure with no inherent timescale can have its timing controlled by the slow viscous deformation of ice. In the simulations, cracks can exist for many years in a stable configuration before very rapidly hydro-fracturing through the full depth, once the slow viscous thinning and bending allows their deepest points to become submerged below the water line. At this point,  a critical stress value is reached and the crevasse propagates abruptly through the whole thickness of the iceberg.  To explain the mechanism for this rapid stage of crevasse propagation, we note that the model assumes that crevasses flood with water once they reach the water level. In terms of the regularised phase-field formulation there is a softer zone ahead of the crack that can contain water and act as an additional stress that propagates the crevasse.

To illuminate the main ideas behind the above failure mechanism, we can consider a highly simplified model. Initial crevasse depth estimates due to \citep{nye_comments_1955,weertman_can_1973} postulate that a crevasse propagates until a depth $d_0$ that is a fraction $R_0$ of the iceberg thickness $h_0$. As noted above, the iceberg thins and bends due to the viscous creep, while the crevasse maintains a fixed vertical extent, so that the crevasse tip advects towards the water line and can fill with water. Once the crevasse occupies a certain higher fraction $R$ of the iceberg thickness $h(t)$ at a time $t$, the crevasse will abruptly propagate towards the base of the iceberg.

Assuming no accumulation and ablation and constant density of the iceberg we can derive a strain thinning approximation in the flow line as outlined in \citep{jansen_model_2005} 
\begin{align*}
\frac{\partial h}{\partial t}  = - C h^4,
\end{align*}
where $C = C(A, \rho_s, \rho_w, \bm{g})$ is a constant dependent on material parameters.\\
The above equation can be solved exactly
\begin{align}\label{eq:thinning}
h(t) = (h_0^{-3}+3Ct)^{-1/3}.
\end{align}
If we require that the initial crevasse depth $d_0$ is equal to $R h(t_{crit})$ at a critical time $t_{crit}$ when the crevasse propagates:
\begin{align*}
Rh(t_{crit}) = d_0 = R_0 h_0,
\end{align*}
from \eqref{eq:thinning} we obtain an estimate on timing of the crevasse propagation
\begin{align}\label{eq:time_crit}
t_{crit} = \frac{\left(\frac{R}{R_0}\right)^{3}-1}{3C} h_0^{-3}.
\end{align}
It is clear that the simplified estimate \eqref{eq:time_crit} cannot accurately capture the exact timing of calving as it neglects flexure near the ice front and other processes. However, the assertion that the time to failure $t_{crit}$ scales with $h_0^{-3}$ seems appropriate, as can be seen in Figure~\ref{fig:timing}(a).

Because our numerical model contains a detailed representation of the fracture process, and because we simulate calving in such a simplified geometrical situation (an iceberg at flotation, with no lateral variations across flow) our simulations can serve as a simple reference case for comparison with other calving laws. 

A number of heuristic calving laws are currently under evaluation to test their suitability as boundary conditions for large scale simulations of ice sheets. Four different laws are compared in \citep{choi_comparison_2018}. These are 1) height above buoyancy, 2) eigencalving, 3) crevasse-depth, and 4) von Mises calving. It is worth briefly considering the qualitative similarities and differences between these four calving laws and the results of our simulations. 

The height above buoyancy criterion qualitatively captures the idea that calving rate increases with ice thickness. However, because the height above buoyancy is zero in our simulations, this law is not appropriate for the floating iceberg that we simulate here: the law would not predict any calving to occur, which is in contradiction to our simulations. 

The eigencalving law considers the calving rate to be proportional to the product of two principal horizontal strain rates. When applied to the situation that we simulate, with no lateral strain rate across flow, this would also predict zero calving rate, so this is also in contradiction to our results. 

The von Mises criterion as implemented by \citet{choi_comparison_2018} adds positive principal strain rates in quadrature, so this law would predict calving to occur in our simulations. Furthermore, because the calving rate is assumed to scale with stress it will increase with ice thickness, just as occurs in our simulations. However, this law also has inconsistencies with our simulations. First, the calving rate is prescribed to increase with stress, and this increases linearly with ice thickness, rather than to the fourth power, as occurs in our results, so this law may seriously underestimate calving rates at large ice thickness. Secondly, the calving rate in most implementations of this scheme is assumed to scale with the ice flow speed. This introduces a problematic frame-dependence to the calving law. Adding a lateral translation to all horizontal velocities would not alter the calving rate in our iceberg simulations, so any explicit dependence of calving-rate upon ice velocity seems problematic. In this law it seems likely that the ice speed is playing a role as a proxy for other variables such as ice thickness. 

Turning to the crevasse-depth criterion developed by \citet{benn_calving_2007} and used in some large-scale ice sheet models \citep{deconto_contribution_2016}, we do find some similarity in behaviour. However, even here, there is a crucial difference. In the \citet{benn_calving_2007} calving model a crevasse at the Nye depth that does not reach the waterline will not produce calving. However, our results show that a very important control on the timing of calving is the time that it takes for the bottom of the crevasse to advect downwards towards the water surface, whereupon sudden hydro-fracture occurs. In fact, in our coupled viscoelastic fracture model, this is the only control on time to failure, since we are modelling brittle failure as a rate independent process, with no inherent timescale, and any elastic adjustments to stress can also happen instantly. Thus, our results support the idea that the crevasse-depth criterion is based on a sound principle, which is that hydrofracture occurs when the deepest part of a crevasse reaches the waterline. However, there is more to the process than this and a full treatment of the advection of crevasses in both horizontal and vertical directions is an important component of the calving problem.

Other calving models have been proposed in addition to those considered by \citet{choi_comparison_2018}. One example is the calving parameterisation advanced by \citet{crawford_marine_2021}. This exhibits a power-law dependence of calving rate on ice thickness, with an exponent in the range 6.0 to 7.3. This is qualitatively similar behaviour to our simulations, but the exponent is higher than the fourth power that is recovered from our simulations. Possible reasons for this difference are that \citet{crawford_marine_2021} consider grounded ice fronts, while we consider floating ice fronts. Other differences are that the workflow used by \citet{crawford_marine_2021} transfers broad-scale geometry from a viscous continuum model to a brittle--elastic discrete-element model, but does not transfer the location of cracks back to the viscous model. Under this workflow, preexisting cracks are not advected downwards towards the waterline by the viscous flow, so the mechanism that we identify as controlling the time to failure would not be reproduced.

In other models, the time to failure is controlled by a timescale inherent in the fracture process itself via a damage evolution equation \citep{krug_combining_2014,mercenier_calving_2018,pralong_dynamic_2005}. Such rate dependence may be important in ice, but it is interesting that our simulations give a plausible explanation for the delay before calving occurs, and show that it is not a necessary condition to have a rate dependence attached to the damage process in order to have a finite time between the calving of successive icebergs.

Another hypothesised mechanism for calving and cliff collapse is failure under compressive shear \citep{bassis_upper_2012,schlemm_simple_2019}. In our simulations, failure can be generated either by a tensile component of deviatoric elastic strain or by tensile volumetric elastic strain. This means that failure under compressive shear can occur in our model. However, Figure~\ref{fig:creep_deformation} also shows that the interaction between the stress field, the fracture network and the water pressure allows tensile forces to develop ahead of the crack tip, eventually promoting full-depth failure.

The above comparisons illuminate some of the similarities and differences between the results from the phase-field model and various other calving laws that have been proposed. These comparisons do not represent a complete validation of the phase-field model and a much fuller comparison with observational datasets will be needed to assess the performance of the model and its ability to capture the rates of calving in more realistic three-dimensional settings found in Antarctica.

\section{Conclusion}\label{sec:conclusions}
Importantly, the model presented here simulates the material behaviour of ice from first principles, but was not constrained to follow any preconceived style of calving. Despite this, it reproduces the commonly observed phenomenon of full-depth block calving: separation of icebergs with horizontal dimension comparable to thickness, leaving a newly-exposed vertical ice front.  We have demonstrated that the system of equations presented here can be used to model a freely floating iceberg, numerically simulating the energetic transfers among gravitational potential energy, stored elastic energy, irrecoverable surface energy and dissipated heat. This is done in a thermodynamically consistent way. 

By representing the lower-dimensional crack network with a regularised phase-field that can be evaluated on the computational mesh, a standard finite-element approach can be used, even for complicated crack networks. The regularisation length provides control on this regularisation, so that realistic behaviour of cracks is recovered when it is sufficiently small relative to the size of the simulated iceberg. The implementation of water pressure within cracks takes full advantage of the phase field representation, which will aid computations of the influence that pressurised water has within more complicated crack topology. 

Although we have simulated very simple geometries, there is no reason that the same system of equations cannot be solved in three-dimensions, taking account of the complicated stress patterns that would be generated by lateral shear, lateral convergence or interactions between multiple cracks. 

Having analysed results from the numerical model and the simple model, we can conclude that Meier’s statement that ``iceberg calving is largely a problem in fracture mechanics coupled to ice dynamics" is indeed a useful way to address the problem of calving.

\section*{Funding Statement}
This publication was supported by PROTECT. This project has received funding from the European Union’s Horizon 2020 research and innovation programme under grant agreement No 869304, PROTECT contribution number XX, and from the NERC National Capability International grant SURface FluxEs In AnTarctica (SURFEIT): NE/X009319/1.

\section*{Competing interests}
The authors report no conflict of interest.

\bibliography{Phase_field_preprint_v5.bib}
\bibliographystyle{jfm}
\newpage

\appendix

\section{Inclusion of pressure inside of cracks}\label{app:phase_field}
In the following section, we derive an approximation that allows the external power due to the hydrostatic pressure boundary condition imposed inside of cracks $\mathcal{C}_i$ to be rewritten as
\begin{align}
\int_{\cup_i \partial \mathcal{C}_i} p_w \mathbf{n} \cdot \dot{\mathbf{u}} \diff S &\approx
\int_{\Omega} 2 p_w (1-d) \nabla d \cdot \dot{\mathbf{u}} \diff V.
\end{align}
Note that $\nabla d$ can be understood as an approximation of the unit normal $\mathbf{n}$ and the term $(1-d)$ ensures that the water pressure $p_w$ acts only inside the softened zone that surrounds the crack. We note that other approaches involving Biot's theory of poro-elasticity \citep{biot_general_1941} are examined in the literature \citep{bourdin_variational_2012,clayton_stress-based_2022,mikelic_phase-field_2015-1,mikelic_quasi-static_2015}.\\

We start with the following expression of the divergence theorem,
\begin{align}\label{eq:pressure_divergence}
\int_{\Omega_{B}} \nabla \cdot (p_w \dot{\mathbf{u}}) \diff V &= \int_{\partial_E \Omega_B} p_w \mathbf{n} \cdot \dot{\mathbf{u}} \diff S + \int_{\cup_i \partial \mathcal{C}_i} p_w \mathbf{n} \cdot \dot{\mathbf{u}} \diff S,
\end{align}
Note that the integral is defined only over the material part of the domain which is time dependent due to potential fracture propagation. To avoid this issue we approximate the volume integral using the degradation function $g(d)$ as follows:
\begin{align*}
\int_{\Omega_{B}} \nabla \cdot (p_w \dot{\mathbf{u}}) \diff V &\approx \int_{\Omega} g(d) \nabla \cdot (p_w \dot{\mathbf{u}}) \diff V \\
&= \int_{\Omega} g(d) \nabla p_w \cdot \dot{\mathbf{u}} \diff V + \int_{\Omega} g(d) p_w \nabla \cdot \dot{\mathbf{u}} \diff V . 
\end{align*}
We then rewrite the second integral using the {Divergence Theorem} and obtain
\begin{align*}
\int_{\Omega_{B}} \nabla \cdot (p_w \dot{\mathbf{u}}) \diff V &\approx
\int_{\Omega} g(d) \nabla p_w \cdot \dot{\mathbf{u}} \diff V - \int_{\Omega} \nabla (g(d) p_w) \cdot \dot{\mathbf{u}} \diff V + \int_{\partial_N \Omega} g(d)p_w \mathbf{n} \cdot \dot{\mathbf{u}} \diff S \\ 
& =  - \int_{\Omega} p_w \nabla g(d) \cdot \dot{\mathbf{u}} \diff V + \int_{\partial_N \Omega} g(d)p_w \mathbf{n} \cdot \dot{\mathbf{u}} \diff S.
\end{align*}
The last integral in \eqref{eq:pressure_divergence} may be degraded in a similar fashion:
\begin{align*}
\int_{\partial_E \Omega_B} p_w \mathbf{n} \cdot \dot{\mathbf{u}} \diff S \approx \int_{\partial_N \Omega} g(d)p_w \mathbf{n} \cdot \dot{\mathbf{u}} \diff S.
\end{align*}
Combining all of the above terms yields an approximation of the hydrostatic pressure boundary condition imposed inside of cracks $\mathcal{C}_i$
\begin{equation}
\begin{split}
\int_{\cup_i \partial \mathcal{C}_i} p_w \mathbf{n} \cdot \dot{\mathbf{u}} \diff S &\approx - \int_{\Omega} p_w \nabla g(d) \cdot \dot{\mathbf{u}} \diff V \\
&= \int_{\Omega} 2 p_w (1-d) \nabla d \cdot \dot{\mathbf{u}} \diff V.
\end{split}
\end{equation}

Note that $\nabla d$ can be understood as an approximation of the unit normal $n$ and the term $(1-d)$ ensures that the water pressure $p_w$ acts only inside of cracks and the surrounding softer area.

\section{Numerical implementation}\label{app:numerics}
In this appendix we define matrices and vectors entering into the finite element iterative solvers from Section~\ref{sec:implementation}. 
We employ the Taylor-Hood elements to ensure numerical stability of the finite element space. The displacements $\mathbf{u}, \mathbf{w}$ are approximated by vectorial piecewise quadratic functions and pressure $p,\ q$ by piecewise linear functions. Fracture phase field $d$ is approximated by piecewise linear functions. \\
At each time step $t_k$ and each iteration we have
\begin{align*}
\mathbf{u} &\simeq \sum_{m} \mathfrak{u}_m \underline{\hat{\varphi}_m}(\mathbf{x}),\\
\mathbf{w} &\simeq \sum_{m} \mathfrak{w}_m \underline{\hat{\varphi}_m}(\mathbf{x}),\\
p &\simeq \sum_{j} \mathfrak{p}_j \varphi_j(\mathbf{x}),\\
q &\simeq \sum_{j} \mathfrak{q}_j \varphi_j(\mathbf{x}),\\
d &\simeq \sum_{j} \mathfrak{d}_j \varphi_j(\mathbf{x}).
\end{align*}
The summation is defined over all degrees of freedom stemming from the finite element discretisation. We define $\varphi_j(\mathbf{x})$ to be the piecewise linear basis functions and $\underline{\hat{\varphi}_m}(\mathbf{x})$ vectorial piecewise quadratic basis functions. Variables denoted in {fraktur} font are the coefficients of the finite element basis functions.

We also need to define projection tensors that allow us to decompose the strain tensors. Recall that a second order tensor $\mathbf{A}$ can be decomposed  
\begin{align*}
\mathbf{A} = \sum_{i = 1}^{3}  \alpha_i \mathbf{m}_i \otimes \mathbf{m}_i,
\end{align*}
where $\alpha_i, \mathbf{m}_i$ are principal strains and principal directions of $\mathbf{A}$.
We define
\begin{align*}
\mathbf{A}_{\pm}:= \sum_{i = 1}^{3}  \langle\alpha_i \rangle_{\pm} \mathbf{m}_i \otimes \mathbf{m}_i,
\end{align*}
where $\langle\ \cdot \rangle_{\pm} = \frac{1}{2}(\cdot \pm |\cdot|)$ is the MacAulay bracket. Note that $\mathbf{A} = \mathbf{A}_+ + \mathbf{A}_-$. The derivative of such decomposition then defines two fourth order projection tensors \citep{miehe_comparison_1998}
\begin{align*}
\mathbb{P}_+ &:= \partial_{\mathbf{A}} \mathbf{A}_+(\mathbf{A}),\\
\mathbb{P}_- &:= \partial_{\mathbf{A}} \mathbf{A}_-(\mathbf{A}) = \mathds{1}- \mathbb{P}_+.
\end{align*}
The projection tensors for a given iteration $(k,i)$ are obtained from the strain tensor from the previous iteration $(k,i-1)$. Therefore, we compute the following fourth order tensors
\begin{align*}
\mathbb{P}_+ &:= \partial_{\mathbf{A}^{(k,i-1)}} \mathbf{A}_+(\mathbf{A}^{(k,i-1)}),\\
\end{align*}
We define the relevant matrices to solve the linear systems of equations \eqref{eq:lin_sys1} and \eqref{eq:lin_sys2}:
\begin{equation}\label{eq:matrices}
\begin{aligned}
\bm{K}^{j,l} &= \int_{\Omega} g(d^{(k,i-1)}) \nabla_s \underline{\hat{\varphi}_j}: \mathbb{P}_+ : \nabla_s \underline{\hat{\varphi}_i} + \nabla_s \underline{\hat{\varphi}_j}: \mathbb{P}_- : \nabla_s \underline{\hat{\varphi}_l} \diff V,\\
\bm{\tilde{K}}^{j,l} &= \int_{\Omega} g(d^{(k,i-1)}) \eta(\Delta t^{-1} (\nabla_s \mathbf{w}^{(k,i-1)}-\nabla_s \mathbf{w}^{(k-1,I)})) \nabla_s \underline{\hat{\varphi}_j} : \nabla_s \underline{\hat{\varphi}_i} \diff V, \\
\bm{V}^{j,l} &= \int_{\Omega} g(d^{(k,i-1)}) \langle -q^{(k,i-1)} \rangle_{+} \varphi_j \nabla \cdot \underline{\hat{\varphi}_l} + \langle -q^{(k,i-1)} \rangle_{-} \varphi_j \nabla \cdot \underline{\hat{\varphi}_l} \diff V,\\
\bm{M}^{j,l} &= \int_{\Omega} \varphi_j \varphi_i \diff V,\\
\tilde{\bm{M}}^{j,l} &= \int_{\Omega} \mathcal{H}^{(k,i)} \varphi_j \varphi_l \diff V,\\
\hat{\bm{K}}^{j,l} &= \int_{\Omega} \nabla \varphi_j \cdot \nabla \varphi_l \diff V.
\end{aligned}
\end{equation}
The external force vectors are similarly given by
\begin{equation}\label{eq:vectors}
\begin{aligned}
    \bm{F}^j &= \int_{\Omega} g(d^{(k,i-1)})\mathbf{f} \cdot \underline{\hat{\varphi}_j} \diff V, \\
\bm{P}^j &= \int_{\Omega} p_w \nabla g(d^{(k,i-1)}) \cdot \underline{\hat{\varphi}_j} \diff V, \\
\bm{T}^j &= \int_{\partial_N \Omega} g(d^{(k,i-1)})\mathbf{t} \cdot \underline{\hat{\varphi}_j} \diff S,\\
\bm{\mathcal{H}}^{j} &= \int_{\Omega} \mathcal{H}^{(k,i)} \varphi_j \diff V.
\end{aligned}
\end{equation}

As suggested in Section~\ref{sec:implementation} we adopt an approach described in \citep{bathe_finite_1975} referred to as material-nonlinearity-only based on the updated Lagrangian formulation, which allows us to distinguish to the body configuration at different times. We assume that the displacement, pressure, and Lagrange multiplier variables at time $t_k$ can be decomposed as
\begin{align*}
    \mathbf{u}^{k} &= \mathbf{u}^{k-1}+ \delta \mathbf{u}^{k},\\
    \mathbf{w}^{k} &= \mathbf{w}^{k-1}+ \delta \mathbf{w}^{k},\\
    p^{k} &= p^{k-1}+\delta p^{k}, \\
    q^{k} &= q^{k-1}+\delta q^{k},
\end{align*}
where the variables $\mathbf{u}^{k-1}, \ \mathbf{w}^{k-1}, \ p^{k-1}, \ q^{k-1}$ are assumed to be known and determine the equilibrium state at time $t_{k-1}$. We rewrite the system \eqref{eq:lin_sys1} to be solved for variable updates $\delta \mathbf{u}^{k},\ \delta \mathbf{w}^{k},\ \delta p^{k}, \ \delta q^{k} $ so that the small displacement assumption in each time step remains valid. 
We introduce a notation for a vector $\bm{S}$ that accounts for the existing stress from the previous time step $t_{k-1}$:
\begin{align*}
    \bm{S} = -\bm{V}^T \mathfrak{p}^{(k-1,I)} + 2\bm{K} (\mathfrak{u}^{(k-1,I)}-\mathfrak{w}^{(k-1,I)}).
\end{align*}

The equivalent system is then given by 
\begin{align*}\label{eq:lin_sys2}
\begin{bmatrix}
2C_2^{-1}/\Delta t\bm{\tilde{K}} +2\bm{K} & -2 \bm{K} & -\bm{V}^T & \bm{V}^T \\
-2 \bm{K} & 2 \bm{K} & \bm{0} & -\bm{V}^T \\
-\bm{V} & \bm{0} & \bm{0} & \bm{0} \\
\bm{V} & -\bm{V} & \bm{0} & -\frac{3(1-2\nu)}{2(1+\nu)} \bm{M} \\
\end{bmatrix}
\begin{bmatrix}
\delta \mathfrak{w}^{(k,i)} \\
\delta \mathfrak{u}^{(k,i)} \\
\delta \mathfrak{q}^{(k,i)} \\
\delta \mathfrak{p}^{(k,i)}
\end{bmatrix}
=
\begin{bmatrix}
\bm{S} + \bm{V}^T \mathfrak{q}^{(k-1,I)} \\
C_1\left( \bm{F} + \bm{P}\right) + \bm{T} - \bm{S}  \\
\bm{0} \\
\bm{0}
\end{bmatrix} 
\end{align*}
The matrices and vectors entering into the discrete system are computed as in \eqref{eq:matrices} and \eqref{eq:vectors} with the difference that the integrals are computed over the domain configuration $\Omega_{k-1}$ at time $t_{k-1}$. As water pressure varies with depth the integrals \eqref{eq:vectors} depend on the the displacement updates which we resolve using an inexact Newton iteration.

\end{document}